\documentstyle[11pt,psfig,aaspp4]{article}

\newcommand{\vv}[1]{\mbox{\boldmath $#1$}}               

\newcommand{\e} {\vv{e}}          
\newcommand{\n} {\vv{n}}          
\newcommand{\E} {\vv{E}}          
\newcommand{\B} {\vv{B}}          
\newcommand{\k} {\vv{k}}          
\newcommand{\om} {\omega}         
                           
\newcommand{\al} {\alpha}                          
                              
\newcommand{\lam}{\lambda}

\def\muinv{\mu^{-1}}
\def\ellip{{\cal P}}

\pssilent

\begin{document}

\title{Photon Splitting in Strongly Magnetized Plasma}

\author{ Tomasz Bulik}

\affil{Nicolaus Copernicus Astronomical Center, 
Bartycka 18, 00-716 Warsaw, Poland}

\begin{abstract} 
{The process of photon splitting, becomes allowed
in the presence of strong magnetic field.
We calculate the influence of magnetized plasma
on the photon splitting absorption coefficient. We
calculate the  refraction coefficients and polarization vectors 
with the inclusion of the vacuum terms at an arbitrary  value of
the magnetic field, and then find photon splitting matrix element
taking into account  the  terms that vanish in vacuum, but 
may be nonzero in the presence of plasma. We find
the photon splitting rate  in plasma with the density typical
for a neutron star  star atmosphere and identify a region of the
parameter  space where the plasma effects are important. } 
\end{abstract}

\section{Introduction}

The possibility of  photon splitting process  in strong
magnetic field around  $B_c = 4.414\times 10^{13}$~Gauss
has been first noted in the early seventies (\cite{adler,bb70}).
Adler (1971) found that only one type of splitting is allowed,
namely $\perp \rightarrow \parallel +\parallel$, where a
perpendicularly polarized photon splits into two parallel
polarized photons.  This calculation took into account the
vacuum dispersion and also the influence of matter with density
that of a pulsar magnetosphere. The normal modes are polarized
linearly in such conditions. The photon splitting absorption
coefficient for this allowed reaction is 
\begin{equation}
\kappa = 0.12 \left( {B\over B_c}\sin\theta\right)^6 
\left({\hbar\omega\over
mc^2}\right)^5\, .
\label{adler}
\end{equation}
Since then a number of authors have derived the matrix elements,
and absorption coefficients for this process using different
approaches (Papanyan  \& Ritus 1972, Stoneham 1979) and also
recently (Baier, Milshtein \& Shaisultanov 1996, Baring \&
Harding 1997). These papers have settled the controversy
sparked by Mentzel, Berg \& Wunnner (1994) who
suggested that the photon splitting rate may actually be a few
orders of magnitude higher than previously thought. 

This exotic process has found  a number of astrophysical
applications. A natural environment where it may play a role is
provided by neutron stars, since a number of them have magnetic
fields in excess of $10^{12}$~Gauss, and about a dozen of radio
pulsars  has spin down fields larger than $10^{13}$~Gauss. it
has been found that photon splitting plays an important role in
the formation of the gamma ray spectrum of PSR~1509-58, where it
inhibits  emission above 1~MeV (\cite{HBG97}). Soft gamma-ray 
repeaters (SGR) form another class of objects where photon
splitting has been suggested to play an important role. There
were three firmly established SGR's with and the  fourth source 
was recently discovered (\cite{Kouv97}). All of them are
characterized by short durations, below a second, and spectra
with cut-offs around  $30$~keV. A number of arguments have been
presented (\cite{TD95}) pointing that these sources are actually
{\em magnetars}, i.e. neutron stars with magnetic fields
reaching $10^{15}$~Gauss, for which the main pool of energy is
the magnetic field.  Baring (1995) suggested that photon
splitting may be responsible for the spectral cutoffs in the
spectra of SGRs, invoking a possibility of photon splitting
cascades. Thompson and Duncan (1995) showed that even in the
absence of cascades photon splitting influences the shape of the
spectra of SGRs.

In  this work we investigate the  process of photon splitting
not only in vacuum but also in the presence of matter with the
density like that in a neutron star atmosphere, when the plasma
dispersion may be important and for some range of propagation 
directions the normal modes are polarized circularly. The
relevance of plasma effects can have a very substantial
influence on the absorption  coefficients. For example, Bulik
and Miller (1997) have shown that plasma effects leads to a
broad absorption like feature  below $\approx 10$~keV for a
large range of SGR emission models. In section 2 we analyze the
polarization of the normal modes  in the strongly magnetized
plasma, in section 3 we calculate the  photon splitting matrix
elements and absorption coefficients, in section 4 we present
the results and we discuss them in section 5.

\section{Polarization of the normal modes, refraction coefficient}

We use the normal mode formalism to describe 
opacities in strong magnetic field.
 The dispersion equation is
\begin{equation}
\vec k \times \left[ \muinv \left( \vec k \times \vec E\right)\right] +
\left({\omega\over c}\right)^2 \epsilon \vec E =0 \, ,
\label{maxwell}
\end{equation}
where $\epsilon$ is the electric permitivity tensor, $\mu$ is
the magnetic permeability tensor and  $\vec E$ is the electric
field of the wave. 
Equation~(\ref{maxwell}) has been discussed by
e.g.\cite{Ginzburg}, and Meszaros (1992). In general
equation~(\ref{maxwell}) has
three solutions: one representing the longitudinal 
plasma oscillations that do not propagate (Meszaros 1992, p.69), 
and two perpendicular representing the electromagnetic waves. 
In the presence of the magnetic field there exist
two nondegenerate wave solutions of equation~(\ref{maxwell}),
and therefore the medium is birefringent. 
We follow the convenient 
method introduced by Gnedin~\&~Pavlov~(1974) for the
case of tenuous plasma to solve equation~(\ref{maxwell}).

The roots of the dispersion equation are
\begin{equation}
n_j = n_I \pm \sqrt{ n_L^2 + n_C^2}\, , \label{disp-root}
\end{equation}
where $j=1,2$ indicates which normal mode is selected. The
refraction coefficients are given by the real part of $n_j$, and
the total absorption coefficients are proportional to the
imaginary part of $n_j$; $\xi_j= (2\omega/c){\rm Im}(n_j)$, and
\begin{equation}
\begin{array}{rcl}
n_I \equiv& 1 + {1\over 4} (\epsilon_{xx} + \epsilon_{yy}\cos^2\theta
+\epsilon_{zz}\sin^2\theta-\epsilon_{xz}\sin 2\theta\\[3mm]
 & -\muinv_{xx} -\muinv_{yy}\cos^2\theta-\muinv_{zz}\sin^2\theta +
\muinv_{xz}\sin 2\theta ) \; ,
\end{array}
\label{nI-def}
\end{equation}
\begin{equation}
\begin{array}{rcl}
n_L \equiv& {1\over 4}(\epsilon_{xx} - \epsilon_{yy}\cos^2\theta
-\epsilon_{zz}\sin^2\theta+\epsilon_{xz}\sin 2\theta\\[3mm]
 & +\muinv_{xx} -\muinv_{yy}\cos^2\theta-\muinv_{zz}\sin^2\theta +
\muinv_{xz}\sin 2\theta ) \; ,
\end{array}\label{nL-def}
\end{equation}
\begin{equation}
n_C \equiv {i\over 2} ({ \epsilon_{xy} \cos\theta 
+\epsilon_{xz}\sin\theta})\, .
\label{nC-def}
\end{equation}
Here $\muinv_{ab}$ is the $ab$ component of the inverse
$\mu$ tensor, and we use the system of coordinates
with the magnetic field along the $z$-axis, and
the wave vector in the $yz$ plane at an angle $\theta$ to the
magnetic field.

We describe the polarization of the normal modes by the position angle
$\chi_j$ between the major axis of the polarization ellipse and the projection
of the magnetic field $\vec B$ on the plane perpendicular to the wave vector
$\vec k$, and the ellipticity $\ellip$, whose modulus is equal to the ratio of
the minor axis to the major axis of the polarization ellipse, and whose sign
determines the direction of rotation of the electric field:
\begin{eqnarray*}
\ellip_j = {r_j -1\over r_j+1}\, , &
r_j\exp(2i\chi_j) = {\displaystyle 
  -n_C \pm \sqrt{ n_L^2 + n_C^2}\over \displaystyle n_L}\,.
\end{eqnarray*}
The polarization vectors can be described by a complex variable
$b$, or by two real parameters $q$ and $p$ (Pavlov, Shibanov, \&
Yakovlev 1980)
\begin{equation}
b  \equiv q+ ip = {n_L\over n_C}\, . \label{qpdef}
\end{equation}
 
Let us consider a wave travelling in the
direction described by  $\theta$ and $\phi$ 
in the coordinate system  with the magnetic field along the
$z$-axis, and the $x$-axis chosen arbitrarily.
The wave vector is $\vec k = k (\cos\theta,
\sin\theta\sin\phi, \sin\theta\cos\phi)$. 
In the rotating  coordinate system
$e_\pm = 2^{-1/2}(e_x \pm i e_y )$, $e_0 =e_z$,
the polarization vectors are
\begin{equation}
\begin{array}{rcl}
e_\pm^j &=& i^{j+1} {1\over\sqrt{2}} C_j e^{\mp i\phi}\left( K_j\cos\theta \pm
1\right)\\[3mm]
e_0^j &= & C_j K_j \sin\theta
\end{array}
\label{vectors}
\end{equation}
where
\begin{equation}
K_j = b \left[ 1 + (-1)^j \left(1 + b^{-2}\right)^{1/2}\right] 
\label{K-def}
\end{equation}
and $C_j = (1+ |K_j|^2)^{-1/2}$. In general $K_j$ are complex.


\subsection{Vacuum  polarization effects}


We  use the convention where we label the polarization
state by the direction of the electric field vector of the 
wave in relation to the external magnetic field. Adler (1971)
used a different convention using the magnetic, 
not electric field of the wave to define polarization.

The refraction indices in strong magnetic field have been derived
by Adler (1971). The result is
\begin{equation}
n_{\parallel,\perp}^{vac} = 1 - {1\over 2}\sin^2\theta A^{\parallel,\perp}
(\omega\sin\theta, B)\, ,
\label{adler-n}
\end{equation}
where the functions $A^{\parallel,\perp}$ are rather lengthy double integrals 
given by equation~(51) in Adler (1971):
\begin{equation}\begin{array}{l}
A^{\parallel,\perp} (\omega, B) =  
{\displaystyle ={\alpha\over 2\pi} 
\int_0^\infty {ds\over s^2} \exp(-s) \int_0^s \exp[\omega^2 R(s,t)]
J^{\parallel,\perp}(s,v)}
\end{array}
\end{equation}
and $v= 2t s^{-1} -1$. The remaining functions $J^{\parallel,\perp}$ are
\begin{equation}
 J^\perp (s,v) = \displaystyle - {Bs \cosh (Bsv)\over \sinh(Bs)} + \\
              \displaystyle + {Bsv \sinh(Bsv)\coth(Bs)\over \sinh(Bs)} +
              \displaystyle - {2Bs[\cosh(Bsv) - cosh Bs)] \over \sinh^3(Bs)}
\end{equation}
\begin{equation}
\displaystyle J^{\parallel} = \displaystyle {Bs\cosh(Bsv)\over \sinh(Bs)}
              \displaystyle -Bs\coth(Bs)\left[ 1- v^2 + v {\sinh(Bsv)\over \sinh(Bs)}
              \right] \, .           
\end{equation} 
The function $R$ is given by 
\begin{equation}
R(s,t) = {1\over 2} \left[ 2t \left( 1- {t\over s} \right) + 
        {\cosh(Bsv) -\cosh(Bs) \over B \sinh(Bs)}\right] \, .
\end{equation}

\begin{figure}
\psfig{file=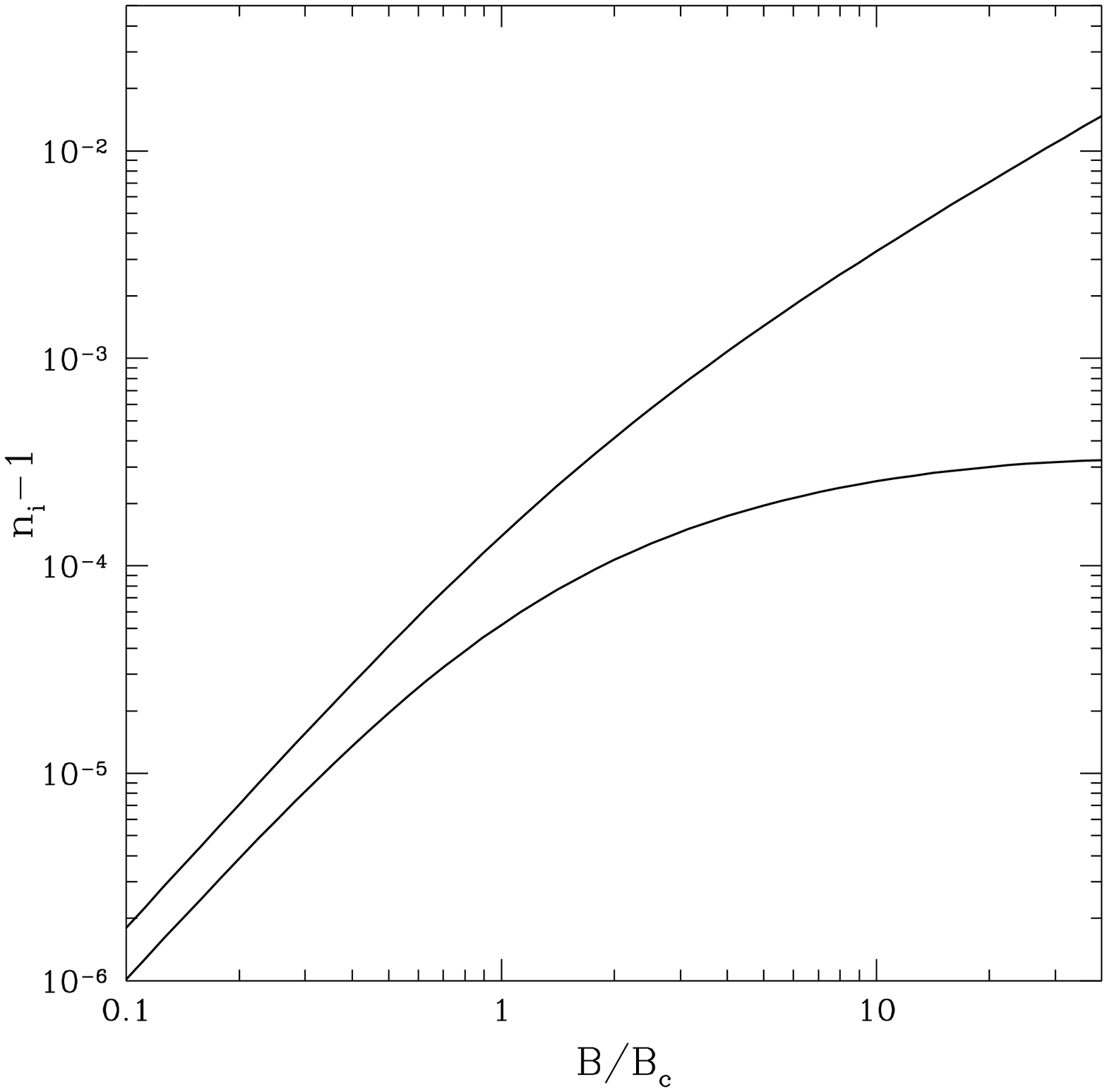,width=0.8\textwidth}
\caption{Refraction coefficients in magnetized vacuum as a
function
of  the 
magnetic field, for propagation angle  of $90^\circ$.
The upper line corresponds to the $\perp$ mode and the lower
curve to the 
$\parallel$ mode. }
\label{n-of-b}
\end{figure}

The refraction indices $n_{1,2}^{vac} = 1+ \eta \sin^2\theta$ 
have been calculated for
$\hbar\omega\ll m_ec^2$
and arbitrary magnetic field by Tsai \& Erber (1974)
 In the low field limit, $B\ll B_c$,
\[
\begin{array}{rcl}
\eta_\parallel(h)&\approx & \displaystyle{14\over 45} h^2 - {13\over 315} h^4  \\[3mm]
\eta_\perp(h)&\approx &\displaystyle {8\over 45} h^2 - {379\over 5040} h^4 \, ,
\end{array}
\]
and when $B\gg B_c$
\[
\begin{array}{rcl}
\eta_\parallel (h)&\approx & \displaystyle 
{2\over 3} h + \left( {1\over 3} + {2\over 3}\gamma 
-8L_1\right) \\[3mm]
\eta_\perp (h) &\approx &\displaystyle {2\over 3} - h^{-1}
\ln(2h)\, ,
\end{array}
\]
where $h=B/B_c$, and $L_1 = 0.249..$, and $\gamma = 0.577..$ is
the Euler's constant (Tsai \& Erber 1974).
Thus in the limit of strong magnetic field refraction for the
parallel mode grows linearly with the field while that of the
perpendicular mode  is nearly  constant. We present the
dependence
of the vacuum refraction coefficients as a function of the
mangetic field in Figure~\ref{n-of-b}.

We have evaluated numerically 
the integrals of equations~(\ref{adler-n}).   
Figure~\ref{n-of-om} shows the dependence
of the refraction coefficients on the photon energy for a few 
magnetic fields.

\subsection{Refraction  coefficient in the system of plasma and vacuum}

Plasma in strong magnetic field 
can be described by the dielectric tensor
\[
\epsilon_{ab} = \epsilon_{ab}^{vac} +\epsilon_{ab}^{p}- \delta_{ab}\, ,
\]
and the vacuum permeability tensor. The plasma dielectric tensor
can be expressed as $\epsilon_{ab} = \displaystyle\delta_{ab} -
\left(\omega_p^2\over \omega^2\right)
\Pi_{ab}$,  where $\omega_p = \displaystyle \left( {4\pi N e^2\over m}
\right)^{1/2} $ is the plasma frequency and $\Pi_{ab}$ is 
the plasma polarization tensor.
The plasma polarization  tensor is diagonal in the rotating coordinates,
and for cold electron plasma   is given by
\begin{equation}
\Pi_{\alpha\alpha}= {\omega\over \omega + \alpha\omega_B -i \gamma_r}
= {\omega \over \omega_t +\alpha\omega_B} \, ,
\label{coldPi}
\end{equation}
where $\alpha=-1,0,+1$, and $\gamma_r= (2/3)(e^2/mc^3)\omega^2$ is
the radiative width, and we denote $\omega_t = \omega -i\gamma_r$.
 Inserting equation~(\ref{coldPi}) into
equations~(\ref{nL-def}) and~(\ref{nC-def}), we obtain
\begin{eqnarray}
n_I &=& 1  - {1\over 4}{\omega_p^2\over \omega} \left[ 
(1+\cos^2\theta) {\omega_t\over \omega_t^2-\omega_B^2}
+ \sin^2\theta {1\over\omega_t}
\right] +  {1\over 4} (A^{\parallel}(\omega \sin\theta, B)+
A^{\perp}(\omega \sin\theta, B)) \, ,
\label{nI-sys}
\end{eqnarray}
\begin{equation}
\begin{array}{l}
\displaystyle n_L = -{\sin^2\theta\over 4} \times 
{\displaystyle\left\{{\omega_p^2 \over \omega\omega_t}
{\omega_B^2\over \omega_t^2-\omega_B^2 } 
+[A^{\parallel}(\omega \sin\theta, B) -
A^{\perp}(\omega \sin\theta, B) ]\right\}}
\label{nL-sys}
\end{array}
\end{equation}
and
\begin{equation}
n_C= -{1\over 2} {\omega_p^2 \over \omega} \left(\omega_B\over
\omega_t^2 - \omega_B^2\right)\cos\theta\, .
\label{nC-sys}
\end{equation}
 Inserting equations~(\ref{nL-sys}) and~(\ref{nC-sys}) into 
equation~(\ref{qpdef}), and neglecting terms proportional 
to $\gamma_r^2$ we obtain  
\begin{equation}
\begin{array}{l}
\displaystyle q =   {\sin^2\theta \over 2\cos\theta}{\omega_B\over\omega}
\displaystyle \left[ 1-(A^{\parallel}(\omega \sin\theta, B) -
A^{\perp}(\omega \sin\theta, B)) 
{\omega^2\over\omega_p^2} \left( {\omega^2\over\omega_B^2} -1\right)
\right]
\end{array}
\label{q-sys}
\end{equation}
and
\begin{equation}
\begin{array}{l}
\displaystyle p= {\sin^2\theta \over 2\cos\theta} \times 
{\displaystyle {\omega_B\gamma_r\over \omega^2}
\left[ 1 + 2  (A^{\parallel}(\omega \sin\theta, B) -
A^{\perp}(\omega \sin\theta, B){\omega^2\over\omega_p^2}
\right] \, .}
\label{p-sys}
\end{array}
\end{equation}
polarization vectors and refraction coefficients are determined
by equations~(\ref{nI-sys}), (\ref{nL-sys}),and (\ref{nC-sys}) 
combined with~(\ref{disp-root}) and (\ref{vectors}). 

The presence of matter influences refraction also because
of the electron cyclotron resonance at $\omega_b$. 
These effects increase for propagation direction along the
magnetic field, and are proportional to the matter density.
For a detailed discussion see e.g. M{\'e}sz{\'a}ros (1992).

\section{Photon splitting rate}

 We will  use two coordinate
systems U, and U'. The z-axis
in the system U lies along the magnetic field, and the
wave vector of the initial photon is in the zx-plane.
The {\em z}-axis in the system U' lies along $\vec k$, the wave vector
of the initial photon,  and
the magnetic field lies in the {\em xz}-plane.
The system U  is convenient for calculation of the matrix element
$M$ and the refraction indices $n_j$, while U' will be used for integration.
In U the wave vector of the initial photon is
$\vec k_0 = k_0(\cos\theta_0,0,\sin\theta_0)$.
The vector $\vec k_1$  in U' is $\vec k_1 = ( \cos\theta',
\sin\theta'\sin\phi', \sin\theta' \cos\phi' )$, and in U 
it is $\vec k_1 = k_1 (
           \cos\theta_0\cos\theta' - \sin\theta_0\sin\theta'\cos\phi',$
            $\sin\theta'\sin\phi',$
             $\sin\theta_0\cos\theta' + \cos\theta_0\sin\theta'\cos\phi'
                )$.
Using the momentum conservation  we obtain in U'
$\vec k_2 = \vec k - \vec k_1$.
Polarization vector  of a photon is given by 
$\vec e(\vec k) = \vec e (\omega,\theta,\varphi)$. We calculate
the polarization vectors for each photon separately, given its 
frequency and the direction of propagation.

The photon splitting 
absorption coefficient can be found by integrating the S-matrix
element over the phase space of the final states,
\begin{equation}
r = \int {1\over 2 }{1\over 2\omega} {d^3k_1\over (2\pi)^3 2 \omega_1}
{d^3k_2\over (2\pi)^3 2 \omega_2} {|S|^2 \over VT}\, .
\label{int-general}
\end{equation}
This  can be expressed as
 \[
 r = {2\alpha^6\over (2\pi)^3} \int |{\cal M}|^2 \omega\omega_1
\omega_2 d^3k_1 d^3k_2 \delta(\sum\omega_i)
\delta( \sum\k_i) \, .
\]
where the  matrix  element $\cal M$ is a function of energies 
and  polarization vectors  of the incoming and outgoing photons.

\subsection{Matrix elements}
 
We use the  system of units in which  $\hbar=c=m_e=1$, 
so the fine structure 
constant is $\al=e^2$. 
The magnetic field is expressed in the units of the
critical field $B_c$.
 The effective Lagrangian of the electromagnetic field
 with QED corrections is given by 
(Berestetskii, Lifshits and Pitaevskii 1982)
\begin{equation}
\displaystyle L_{\rm eff} = 
\frac{1}{8\pi^2} \int_0^\infty 
\frac{e^{-\lam} \, {\rm d} \lam}{\lam^3} 
{\displaystyle\times\left\{ -(\xi \cot \xi) (\eta \coth \eta) + 1 - 
       \frac{\xi^2 - \eta^2}{3} \right\}\, .}
\end{equation}
Here \begin{eqnarray*}
\xi &=& -\lam\frac{ie}{\sqrt{2}} \left\{ 
      ( {\cal F} + i {\cal J} )^{1/2} - ( {\cal F} - i {\cal J} )^{1/2} 
                           \right\}, 
  \\
\eta &=& \lam \frac{e}{\sqrt{2}} \left\{ 
      ( {\cal F} + i {\cal J} )^{1/2} + ( {\cal F} - i {\cal J} )^{1/2} 
                        \right\} ,
\end{eqnarray*}
and $
{\cal F} = \frac{1}{2} \left( \B^2 - \E^2 \right), \;
{\cal J} = \B \E $
are the field invariants. In the low frequency approximation
the $S$-matrix  element can be calculated as
\[
S_{\gamma \rightarrow \gamma_1 + \gamma_2} = 
\langle \gamma \, 
| \int {\rm d} \vv{r} \, {\rm d} t \, V_{\rm int} | \, 
\gamma_1 \, \gamma_2 \rangle ,
\]
where the interaction operator is $V_{\rm int} = L_{\rm eff}$.

We find that the two lowest order terms of the S-matrix 
correspond to a square and hexagon diagrams:
\begin{equation}
 S_6  = 
-i \frac{ 2 \alpha^3 B^3 }{(2\pi)^2 315 } (4\pi)^{3/2} \times 
{ {  (2\pi)^4 \,
\om \om_1 \om_2 \, 
\delta( \sum\om_i) \,
\delta( \sum\k_i ) \, M_6 .}}
\end{equation}
and
\begin{equation}
S_4= i {\displaystyle2\alpha^2\over \displaystyle 45(4\pi)^2}
(4\pi)^{3/2}B \omega\omega_1 \omega_2 \delta( \sum\om_i) \,
\delta( \sum\k_i )\ M_4\, . 
\end{equation}
 The matrix elements are
\begin{eqnarray*}
M_6 & = & 48 {\cal A} +26{\cal B} + 13{\cal C} + 16\cal{D} \, ,\\
M_4 & = & 8{\cal C} + 14 {\cal D}\, ,
\end{eqnarray*}
and denoting the by $n_i$, $\n_i$, $\e_i$, the index of
refraction, the direction of propagation, and the polariztion
vector of th $i-$th photon respectively we obtain:
\[
{\cal A} =n_0 n_1 n_2 
       (\n_0 \times \e_0)_z (\n_1 \times \e_1^\ast)_z (\n_2 \times \e_2^\ast)_z 
\]
\begin{eqnarray*}
{\cal B}& =&n_0 (\n_0\times \e_0)_z e_{1z}^\ast e_{2z}^\ast +
  n_1 (\n_1 \times \e_1^\ast)_z e_{0z} e_{2z}^\ast +
 n_2 (\n_2 \times \e_2^\ast)_z e_{0z} e_{1z}^\ast .
\end{eqnarray*}
\begin{eqnarray*}
{\cal C} & = & n_0 (\n_0 \times \e_0)_z 
       \{ [ n_1n_2 (\n_1\n_2) - 1 ] (\e_1^\ast \e_2^\ast)  - 
          n_1n_2 (\n_1 \e_2^\ast) (\n_2 \e_1^\ast) \} + 
\nonumber \\
&\;\;&   n_1 (\n_1 \times \e_1^\ast)_z 
       \{ [ n_0 n_2 (\n_0 \n_2) - 1 ] (\e_0 \e_2^\ast) - 
          n_0 n_2 (\n_0 \e_2^\ast) (\n_2 \e_0) \} +
\nonumber \\
&\;\;&    n_2 (\n_2 \times \e_2^\ast)_z 
       \{ [ n_0n_1 (\n_0 \n_2) - 1 ] (\e_0 \e_1^\ast) - 
          n_0n_1 (\n_0 \e_1^\ast) (\n_1 \e_0) \}  
\end{eqnarray*}
and
\begin{eqnarray*}
{\cal D}&=&e_{0z} (n_1\n_1 - n_2\n_2 ) (\e_1^\ast \times \e_2^\ast) +
   e_{1z}^\ast (n_0\n_0 - n_2\n_2 ) (\e_0 \times \e_2^\ast) +
   e_{2z}^\ast (n_0\n_0 - n_1\n_1 ) (\e_0 \times \e_1^\ast)
\end{eqnarray*}
  The matrix element  can be expressed as
 \begin{equation}
S= -i {2\alpha^3 \over (2\pi)^2} (4\pi)^{3/2} \omega\omega_1 \omega_2
{\cal M} (2\pi)^4  \delta (\sum\omega_i)
\delta(\sum\k_i)
\label{smatrix}
\end{equation}
where 
\begin{equation}
{\cal M} =  {B^3\over 315} M_6 + {B\over 45 \alpha^2} M_4\, .
\label{m-element}
\end{equation}

\subsection{Polarization selection rules}

In the vacuum, polarization of the normal modes is linear, i.e.:
 \begin{eqnarray}
\n &=& (\cos\theta, 0, \sin\theta) \nonumber \\
\e_\parallel & = & (\sin\theta, 0 ,-\cos\theta) \label{linpol}\\
\e_\perp &=&  (0,1,0)\, .\nonumber 
\end{eqnarray} 
Functions ${\cal C}=0$   and   $\cal D$
vanish when we neglect refraction, while otherwise
they are of the order of $n -1$.
Using equations (\ref{linpol}) and neglecting refraction we find that 
$M_4$ vanishes while  
\begin{eqnarray*}
  M_6 (\perp \rightarrow \perp +\perp) & =& 48 \sin^3\theta \nonumber \\
  M_6 (\perp \rightarrow \parallel + \parallel ) & =& 26 \sin^3\theta \\
  M_6 (\parallel \rightarrow \perp + \parallel ) & =& 26 \sin^3\theta  \nonumber
\, ,
\end{eqnarray*}
and $  M_6 = 0$ for all  other transitions.

In the presence of matter the functions $\cal C$ and $\cal D$ no
longer vanish. This is due to the fact that, in general,  the
refraction coefficients differ from unity, polarization vectors 
are elliptical rather than linear, photons in the process are
not exactly collinear, and  there exists a small  degree of
longitudinal polarization.  The polarization related effects are
the strongest for high matter density and propagation direction
along the field.


\subsection{Kinematic selection rules}

\begin{figure}
\psfig{file=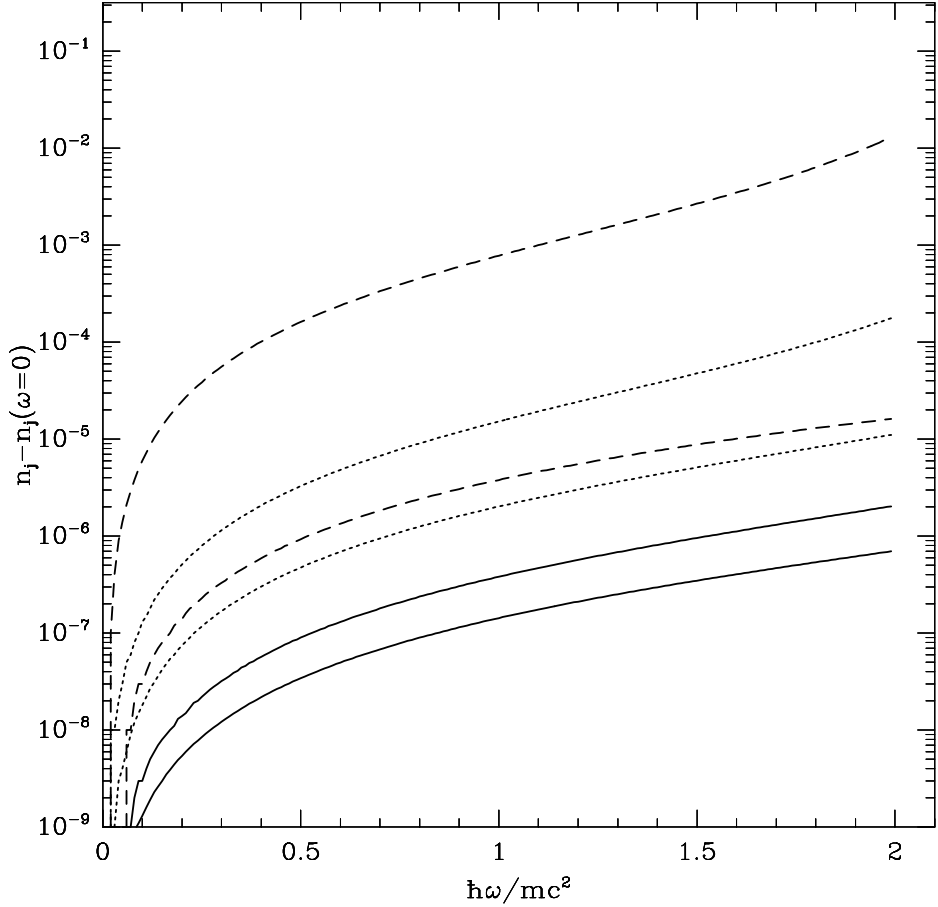,width=0.8\textwidth}
\caption{Refraction coefficients in magnetized vacuum as a
function
of photon energy for three values of the magnetic field: $B=0.3
\times B_c$ 
(solid lines),
$B=B_c$ (doted lines), and for $B=10\times B_c$ (dashed lines).
For each
magnetic
field
the upper line corresponds to the $\perp$ mode and the lower
curve to the 
$\parallel$ mode. The direction of propagation is perpendicular
to the magnetic
field. Note that at photon energies
above $m_e C^2$ the low energy approximation $n-n(\omega=0)
\propto \omega^2$
no longer holds, especially for the very strong field case.}
\label{n-of-om}
\end{figure}

The kinematic selection rules arise from the fact that 
for some transitions the  the energy conservation 
cannot be satisfied. 
The kinematic selection rules can be discussed analytically
using the dispersion relation  $k=n(\omega,\mu) \omega\equiv (1+\xi(\omega,\mu))\omega$. 
We note that the refraction coefficient is very close to unity, 
see Figures~\ref{n-of-b} and \ref{n-of-om},
 and therefore for a given $\mu$ and one can
invert the dispersion relation  to obtain
\begin{equation}
\omega = (1- \xi(k,\mu) +0(\xi^2) ) k 
\label{om-of-k}
\end{equation} 
We note that  for the fixed initial photon the function 
$f \equiv \omega_0 -\omega_1 -\omega_2 $ 
is monotonically decreasing with $\mu'$- the cosine of the angle between
the initial and one of the final photons.
Thus, it suffices to verify that  equation $f=0$ 
has no  solution for the collinear photons, to be sure
that there are no solutions for non-collinear photons. 
Using  momentum conservation and equation~(\ref{om-of-k}),  one
obtains for the collinear photons:
\begin{equation}
f_{\rm collinear} = -\xi(\k_0,\mu_0) k_0 +\xi(k_1,\mu_0)k_1
+\xi(k_2,\mu_0)k_2
\label{f-check}
\end{equation}
 The condition for the kinematic selection rules to be satisfied is
$f_{\rm collinear} > 0$. 

In vacuum in the limit ${\omega\over m} \ll 1$ the refraction
coefficients  are functions of the magnetic field and the
dependence on photon energy $\omega$ is weak, see
Figure~\ref{n-of-om}. In this case condition (\ref{f-check}) can
be rewritten as $-\xi_0+ x\xi_1 +(1-x)\xi_2 > 0$, where $x$ is a
number between $0$ and $1$. It is clear, that transitions
$\parallel \rightarrow \parallel + \perp$  and $\parallel
\rightarrow \perp + \perp  $ are not allowed. Since in the first
order in $\omega\over m$  the functions $\xi$ increase as a
function of $\omega$ transitions $\parallel \rightarrow
\parallel + \parallel$ and  $\perp \rightarrow \perp + \perp$
are also forbidden. Thus in vacuum only two transitions are
kinematically allowed: $\perp \rightarrow \perp + \parallel $
and  $\perp \rightarrow \parallel + \parallel$. 

Combining the polarization selection rules and the kinematic
selection rules  we conclude, in agreement with Adler, that only
the transition $\perp \rightarrow \parallel + \parallel$ is
allowed  vacuum.

In the presence of matter the discussion of kinematic selection
rules becomes more complicated since the refraction indices as a
function of energy  are rather complicated functions.  The
kinematic selection rules become complicated when the field is
strong enough so that the electron cyclotron resonance
influences refraction significantly. Around the resonance
refraction coefficients are non monotonic functions of energy.
In this case there may be a fraction of final state space that
becomes kinematically allowed because of the influence of 
plasma on refraction coefficients.

\subsection{Absorption coefficient}

In the case of propagation in vacuum we ignore  dispersion and  calculate the 
integral of equation~(\ref{int-general})
 \begin{eqnarray*}
\int dr =  {2\alpha^6\over (2\pi)^3} |{\cal M}|^2 \times 
 \int k k_1 (k-k_1) k_1^2 dk_1 d\cos\theta   d\phi  
 \delta(k -k_1 -|\vec k-\vec k_1|) 
 = {\alpha^ 6 \over 2 \pi^2} |{\cal M}|^2 {\omega^5 \over 30}\, .
\end{eqnarray*}
Due to the selection rules discussed above  the only non-vanishing  photon
splitting absorption coefficients is
\[
 r(\perp \rightarrow  \parallel + \parallel ) = {\alpha^6\over 2\pi^2}
B^6 \sin^6\theta {\omega^5\over 30} \left({26 \over 315}\right)^2 
 \, ,
\]
in full agreement with the  result  obtained by Adler (1971).

In general, when the polarization modes are not linear and  we take into
account the dispersion relation equation~(\ref{int-general}) can be written as
\begin{equation} 
\displaystyle r = {2\alpha^6 \over (2\pi)^3} 
{\int |{\cal M}|^2 \omega\omega_1
\omega_2 k_1^2  dk_1 d\cos\theta' d\phi' \delta\left( \omega-\omega_1-\omega_2
\right) }
\label{toiowo} 
\end{equation}
where  we wrote explicitly the variables in the system U' and  $\omega$-s are
function of $k$ and $\mu$.  Defining $f\equiv \omega_0 -\omega_1 -\omega_2$, we
can integrate over $\cos\theta'$ and obtain
\begin{equation}
 r = {2\alpha^6 \over (2\pi)^3}
\int_0^k dk_1   k_1^2 
\int_0^{2\pi} d\phi'\omega \omega_1 \omega_2 { |{\cal M}|^2 }
\left|{df \over d\mu}\right|^{-1}_{f=0} \, ,
\label{int-final}
\end{equation}
which can be evaluated numerically. 
The integrand in equation~(\ref{int-final}) is understood to vanish 
whenever there is no solution of equation $f=0$.

\section{Results}

We evaluate equation~(\ref{int-final}) numerically to obtain the
photon splitting absorption coefficient for different polarization
channels, and various photon energies, magnetic fields and
plasma densities. In the calculation we use the refraction
indices of  equation~(\ref{disp-root}), the polarization vectors
given by equation~(\ref{vectors}), and calculate the matrix
element using equation~(\ref{m-element}).  At each integration
point we evaluate the function $f$ and find whether the energy
conservation is satisfied.

\begin{figure*}
\plotone{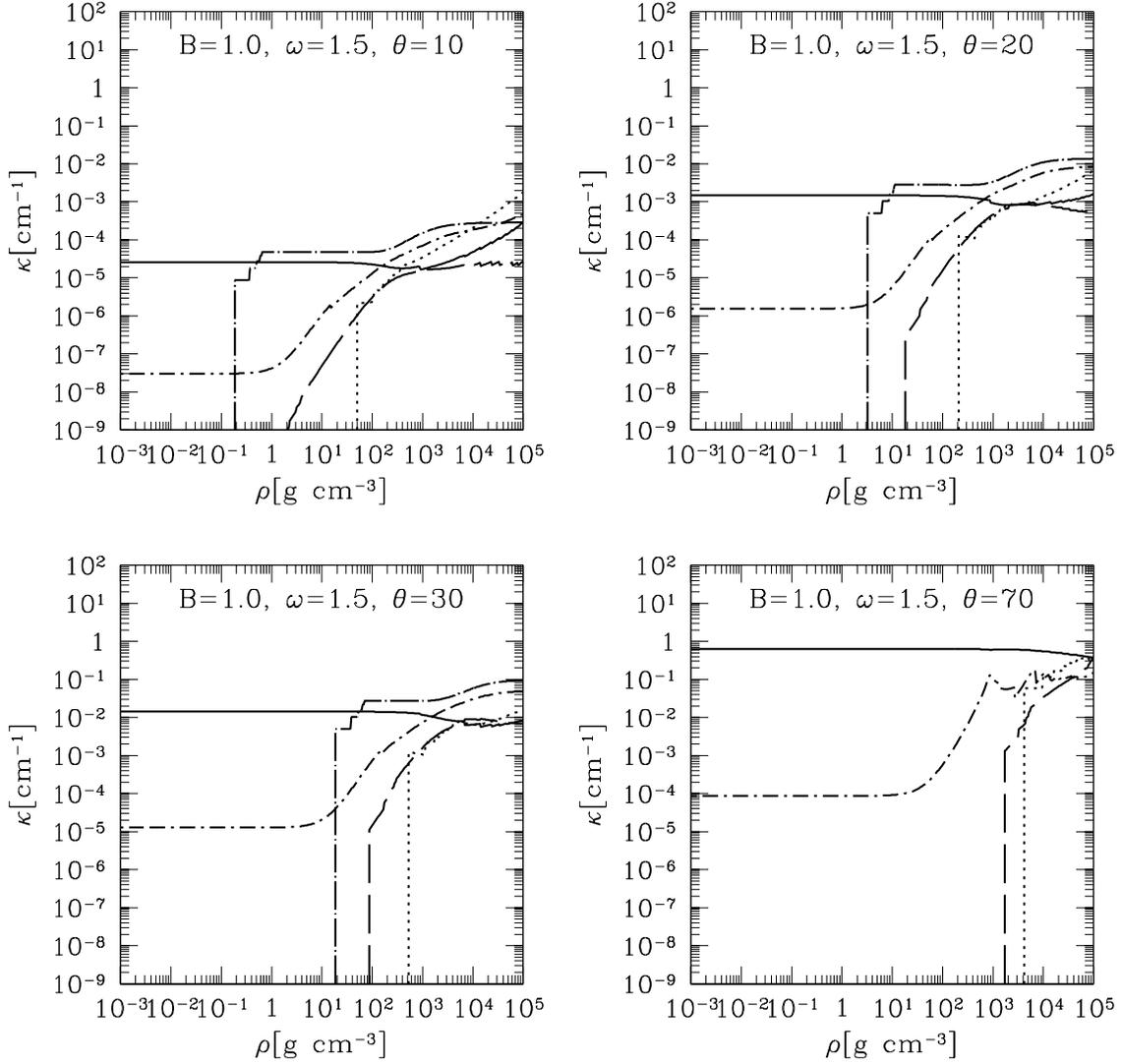}
\caption{Photon splitting absorption coefficient as a function of
density
in plasma for $B=B_c$. The solid line corresponds to 
the case $1\rightarrow 2+2$,
i.e. the transition allowed in vacuum, the dotted line represents
the transition $2\rightarrow 1+2$,  the long dashed $2\rightarrow 2+2$,
the short dashed dotted $1\rightarrow 2+1$,and the long dashed
dotted $1 \rightarrow 1+1$
}
\label{rhob1}
\end{figure*}
\begin{figure*}
\plotone{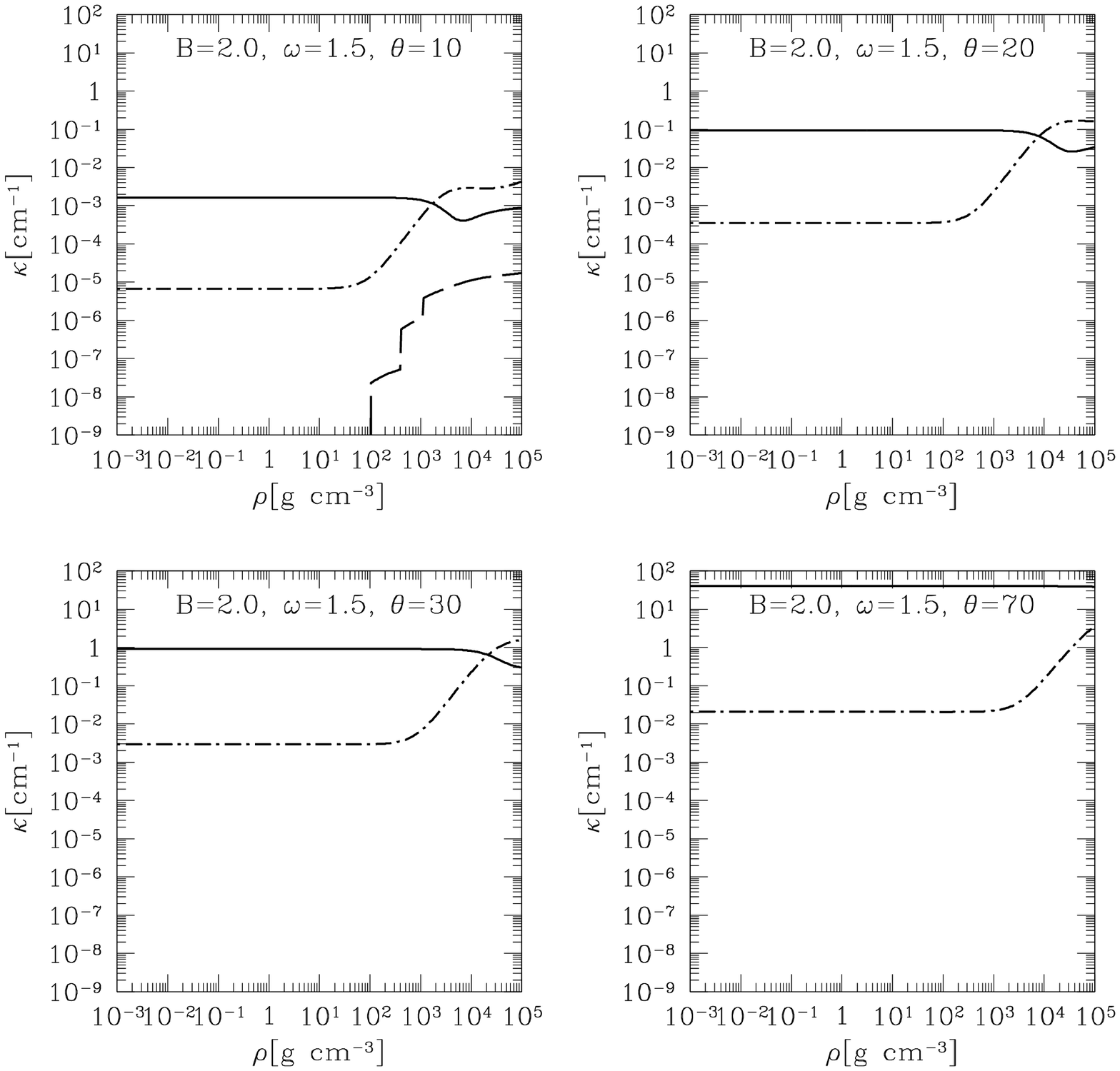}
\caption{Photon splitting absorption coefficient as a function of
density
in plasma for $B=2\, B_c$. For
explanation see Figure 3.}
\label{rhob2}
\end{figure*}
\begin{figure*}
\plotone{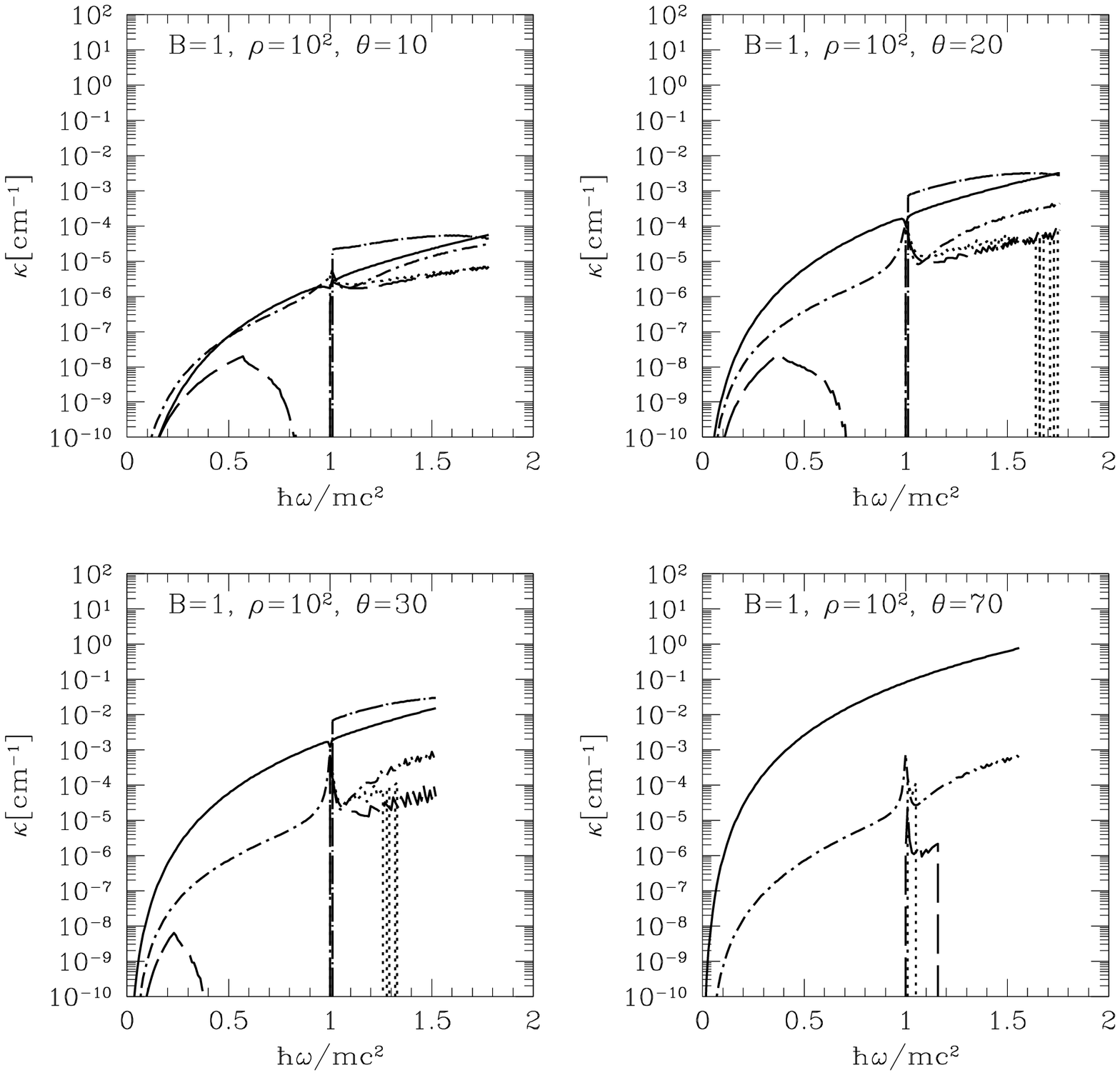}
\caption{Photon splitting absorption coefficient as a function of
energy
in plasma with density 100~g~cm$^{-3}$ for $B=B_c$.For
explanation see Figure 3.}
\label{omb1}
\end{figure*}
\begin{figure*}
\plotone{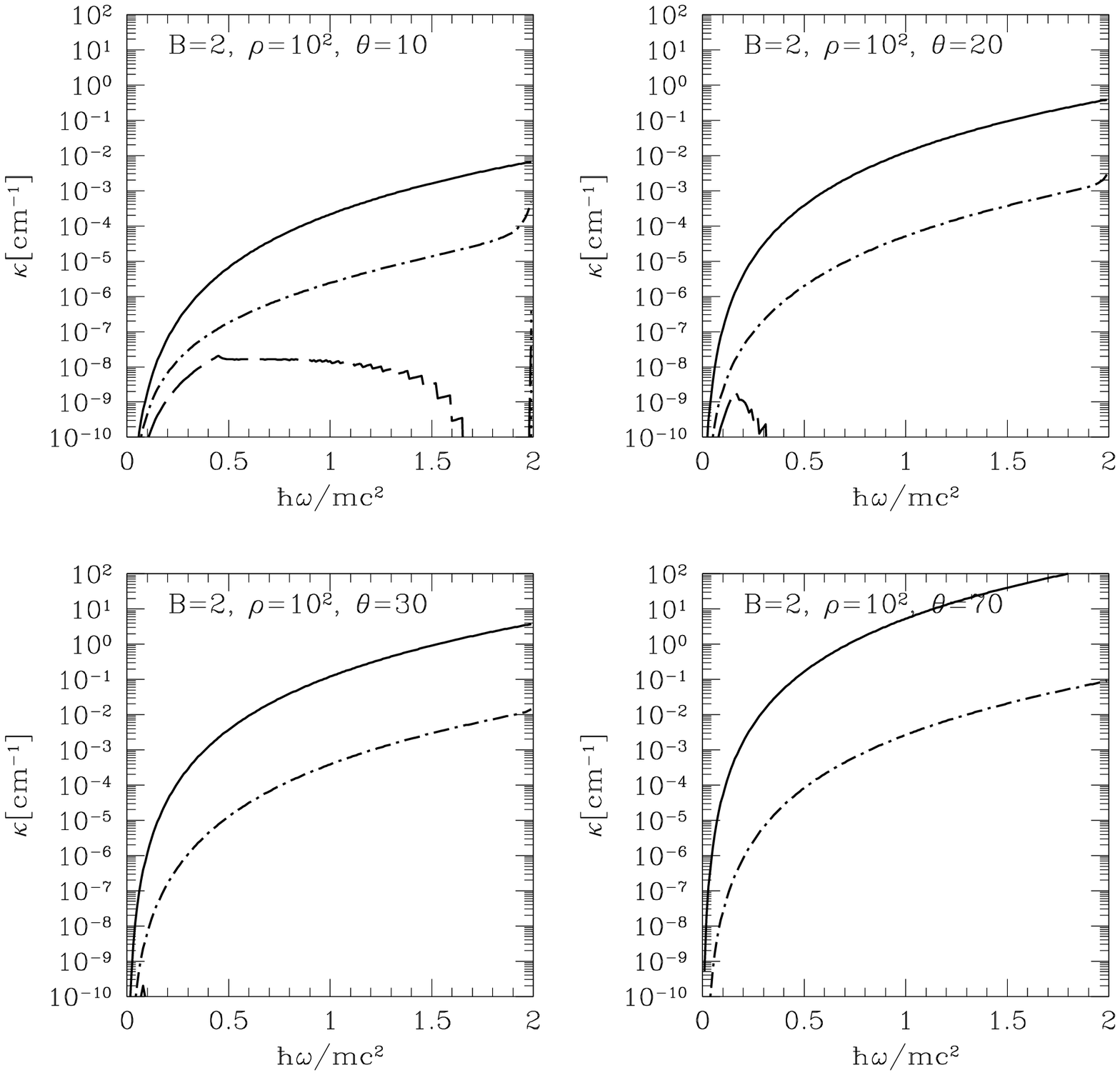}
\caption{Photon splitting absorption coefficient as a function of
energy
in plasma with density 100~g~cm$^{-3}$ for $B=2\, B_c$. For
explanation see Figure 3.}
\label{omb2}
\end{figure*}

We present the results in Figures \ref{rhob1}, \ref{rhob2},
\ref{omb1}, and \ref{omb2}. Each figure consists of four panels
which show the splitting rates for four angles of propagation
with respect to the magnetic field: $10^\circ$, $20^\circ$,
$30^\circ$, and $70^\circ$.  Figures \ref{rhob1}, \ref{rhob2} show the
splitting absorption coefficient as  a function of matter
density of a photon  with the energy $\hbar\omega= 1.5 m_e c^2$;
Figure  \ref{rhob1} for the case $B=B_c$ and Figure  \ref{rhob2}
for the case $B= 2 B_c$. We present photon splitting absorption
coefficients as functions of energy  when the plasma density is
$100$g~cm$^{-3}$ in  for $B=B_c$ in Figure \ref{omb1}, and for
$B=B_c$ in Figure \ref{omb2}. We show the photon splitting
absorption coeficients  for the energies below $2m_ec^2$ since
above this energy the  opacity is dominated by the single photon
pair production.

We first consider the effects of matter density on the photon
splitting rates.   At low densities the influence of matter is
negligible, and we recover the vacuum case when all  the
polarization vectors are linear, and the kinematic selection
rules are  determined by the magnetic vacuum refraction.  With
the increase of density polarization  of the normal modes
becomes  elliptical starting at the photons propagating near to
the direction of the field.  At small angles this effects are
pronounced already  at the density of $0.1$g~cm$^{-3}$. However
comparing Figures 4 and 5 we see that the dominant effect is due
to the kinematic selection rules and the influence of  the
electron cyclotron resonance is crucial. This is also clearly
seen in Figure 6, where a number of  splitting channels that are
forbidden in  vacuum suddenly turns on for energies above the 
cyclotron resonance.  

At a low value of the magnetic field  the process rapidly
becomes unimportant, since the rate scales $B^6$. With the
increasing value of the magnetic field the plasma effects start
to be important when  the electron cyclotron resonance falls
right around the electron mass, i.e. the field has a value close
to the critical field. In these case when the matter is
sufficiently dense the effects of the electron cyclotron
resonance  influence the kinematic selection rules
significantly  thus allowing more photon splitting polarization
channels. When the value of the magnetic field is higher than $2
B_c$ the electron cyclotron resonance falls above  region of
integration over the final states and  the does not influence
the splitting rate. Moreover with the increasing magnetic field
refraction becomes dominated by the vacuum terms, and the
effects of plasma become less and less important.

Figures \ref{omb1} and \ref{omb2} show the effects of photon
energy on the splitting absorption coefficient. When the
electron cyclotron resonance effects are ignored only  small
frequencies and small angles of propagation are influenced,
since it is there where the refraction coefficients and the
polarization of the normal modes are the most influenced by
plasma.  However this is also the region where the photon
splitting coefficient is the smallest, see
equation~(\ref{adler}).

\section{Discussion}

In this work we extend the results of Adler (1971) to the case
of propagation in the magnetized plasma  and concentrate on the
case of  the density typical for a neutron star atmosphere.  Our
approach is accurate for the magnetic  fields up to
approximately the critical field since we use a low field 
approximation  in the calculation of the matrix element.  We
calculate the refraction coefficients accurately and thus the
kinematic selection rules do not  suffer from this limitation. 

We have calculated the photon splitting absorption coefficient
as a function of magnetic field, plasma density, and the photon
energy and direction.  We have found a region of the parameter
space (density $\rho > 1\,$g~cm$^{-3}$, the magnetic field $0.1
B_c < B < 2 B_c$, propagation angles $\theta < 30^\circ$) where
the effects of plasma are the most pronounced. A part of these
region is where the photon splitting absorption is small,  so
the region of importance is limited to photon energies  above
$m_e c^2$. We find that the the photon splitting rate is well
described by the vacuum approximation in the remaining part of
the parameter space.

Photon splitting absorption coefficient  is small when compared
to other processes that may play a role in plasma, for example
the electron scattering opacity is $k_{scat} = 0.4
(\hbar\omega/m_e c^2)^2 (B_c/B)^2 (\rho/$g~cm$^{-3})$cm$^{-1}$  for the 
extraordinary mode, a value a few orders of magnitude  higher
than that for photon splitting. Thus photon splitting can play a
significant role only in very special astrophysical cases.  An
example of such environment  could be in deep layers of a
neutron star atmosphere. Therefore the results presented here may
apply to soft gamma-ray repeaters, where a large amount of
energy is deposited in the crust of a neutron star. Photon
splitting in a high density plasma may be a way of producing a
large number of soft X-ray photons which later on escape. Our
results can also be applied to  the high energy radiation from
isolated neutron stars provided that  we see radiation from the
surface and not the magnetopshere. However, the main conclusion is
that surprisingly the effects due to the presence of plasma are
important in a rather small fraction of the parameter space and
the vacuum approximation can be used in most calculations.

Acknowledgements. This work was supported by the following
grants KBN-2P03D00911, NASA NAG 5-4509 and NASA NAG 5-2868.  The
author thanks Victor  Bezchastnov for assitance in calculating
the matrix elements, and  George Pavlov, Don Lamb  and Cole
Miller for many helpful discussions  during this work.

\end{document}